\documentclass[journal]{IEEEtran}
\ifCLASSINFOpdf
\else
\fi
\usepackage{amsmath,amssymb,amsfonts}
\usepackage{cite}

\usepackage{algorithmic}
\usepackage{stfloats}
\usepackage{graphicx}
\usepackage{subfigure}
\usepackage{textcomp}
\usepackage{xcolor}
\usepackage{pifont}
\usepackage{amsthm}
\newtheorem{theorem}{Theorem}

\usepackage{mathrsfs}
\usepackage{cases}
\usepackage{comment}
\usepackage{empheq} 
\usepackage{caption}
\usepackage{bm}
\allowdisplaybreaks
\usepackage{xcolor}
\usepackage{url}  
\usepackage{graphicx}  
\usepackage[ruled,vlined,linesnumbered]{algorithm2e}
\usepackage{setspace}
\usepackage{floatpag} 
\usepackage{float}
\floatpagestyle{empty}

\renewcommand{\baselinestretch}{0.89}

\usepackage{listings}
\usepackage{fancyhdr} 
\usepackage{url}
\begin{document}
\title{Secrecy Rate Maximization for Reconfigurable Intelligent Surface Aided Millimeter Wave System with Low-resolution DACs}
%
%
%

\author{Yue Xiu,~Jun Zhao,~\IEEEmembership{Member,~IEEE},
~Wei Sun,~\IEEEmembership{Student Member,~IEEE},~Zhongpei Zhang,~\IEEEmembership{Member,~IEEE}
\thanks{Yue Xiu, and Zhongpei Zhang are with the University of Electronic Science and Technology of China, Chengdu, China (E-mail: xiuyue@std.uestc.edu.cn).
Jun Zhao is with Nanyang Technological University, Singapore (E-mail: junzhao@ntu.edu.sg).
Wei Sun is with the School of Computer Science and Engineering, Northeastern University, Shenyang 110819, China (E-mail: weisun@stumail.neu.edu.cn).

This work was supported by 1) Guangdong province Key Project of science and Technology(2018B010115001), 2) Major Project of the Ministry of Industry and Information Technology of China under Grant (TC190A3WZ-2), Nanyang Technological University (NTU) Startup Grant, 3) Alibaba-NTU Singapore Joint Research Institute (JRI), 4) Singapore National Research Foundation (NRF) under its Strategic Capability Research Centres Funding Initiative: Strategic Centre for Research in Privacy-Preserving Technologies \& Systems (SCRIPTS).
}}

\maketitle
 \thispagestyle{fancy}
\pagestyle{fancy}
\lhead{This paper appears in \textbf{IEEE Communications Letters}. \hfill \thepage \\ Please feel free to contact us for questions or remarks. }
\cfoot{}
\begin{abstract}
In this letter, we investigate the secrecy rate of a reconfigurable intelligent surface (RIS)-aided millimeter-wave (mmWave) system with low-resolution digital-to-analog converters (LDACs). Compared to the RIS-aided systems in most existing works, we consider how to alleviate the hardware loss and improve the secrecy rate by using RIS. In particular, we formulate a secrecy rate maximization problem with hardware constraints. Then, the RIS phase shifts and the transmit beamforming are optimized to maximize the secrecy rate.  Due to the nonconvexity of the problem, the formulated problem is intractable.  To handle the problem, an alternating optimization (AO)-based algorithm is proposed. Specifically,  we first use the successive convex approximation (SCA) method to obtain the transmit beamforming. Then, the element-wise block coordinate descent (BCD) method is used to obtain the RIS phase shifts.  Numerical results demonstrate that the  RIS can mitigate the hardware loss, and the proposed AO-based algorithm with low complexity outperformances the baselines.
\end{abstract}


%
\IEEEpeerreviewmaketitle

\section{Introduction}
Millimeter-wave (mmWave) technologies have played an important role in 5G communication systems. Compared to the microwave, the mmWave can achieve a large system capacity and security performance \cite{b1}. However, the blockage issue needs to be tackled before the commercial application of mmWave technologies. Specifically, mmWave frequencies are susceptible to blockage, which means mmWave communications are difficult to be applied in urban areas with dense buildings \cite{b2}.

To handle this problem, reconfigurable intelligent surface (RIS) is proposed as a promising technology \cite{bc1,bc2,bc3,bc4,b5}. In particular, by adjusting the reflection matrix of RIS, the propagation direction of the transmitted signal can be changed, mitigating mmWave communication systems' blockage problem. {Therefore, RIS-aided mmWave systems are getting growing interest from both academia and industry \cite{b8,b9}.}

{Physical layer security is a key technology for solving privacy protection problems in the physical layer and has been a hot topic in the past decade.  To improve security performance, the basic idea is to improve legitimate users' achievable rate or deteriorate that of the eavesdroppers (Eve). In a RIS-aided wireless system, the received signal can be suppressed at Eve while being boosted at the legitimate user. Thus, deploying the RIS bring a new degree of freedom (DoF) in the space domain, and the secrecy rate can be further improved.}

Recently, the secure problem of RIS-aided mmWave systems has been investigated in \cite{b20,b21,b22}. Specifically, in \cite{b20}, the authors maximized RIS-aided communication systems' secrecy rate by optimizing the RIS phase shifts and the transmit beamforming. In \cite{b21}, a low-complexity iterative algorithm was proposed to solve the sum secrecy rate maximization problem for RIS-aided multi-user mmWave system. In \cite{b22}, the authors optimized the hybrid precoding at APs and phase shifting at the RIS to maximize RIS-aided mmWave systems' secrecy rate.

However, the RIS-aided mmWave security systems with hardware limitations have not been investigated yet. In general, since much more power consumption is needed for the high-resolution digital-to-analog converts (HDACs) \cite{b10}, to cut the hardware cost and power budget, low-resolution digital-to-analog converts (LDACs) have been widely used in the mmWave system. However, the hardware imperfections of LDACs and RIS phase noise caused by finite discrete phase shifts of the RIS are another crucial hardware impairment in RIS-aided systems \cite{b6,b13}.  Hence, we focus on a RIS-aided mmWave security system with LDACs and phase noise in this paper. Specifically, we maximize the secrecy rate under these hardware constraints by jointly optimizing the transmit beamforming and the RIS phase shifts. Since the objective function and the feasible set are non-convex,  the formulated problem is intractable. To cope with these difficulties, we propose an alternating optimization (AO)-based algorithm based on the successive convex approximation  (SCA)  method and the element-wise block coordinate descent (BCD)  method. 

\section{System Model}
\begin{figure}[!t]
\centering
\includegraphics[scale=0.45]{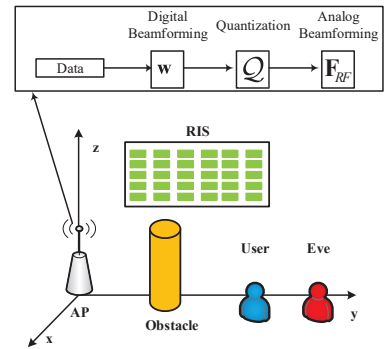}
\caption{A RIS-aided mmWave secure communication system with hardware limitations.\vspace{-20pt}}
\end{figure}
We consider a RIS-aided massive MIMO mmWave downlink, as shown in Fig.~1, where the AP with $N_{t}$ antennas and $N_{RF}$ RF chains sends data to the user and Eve. The user and Eve are equipped with a single antenna, respectively. Considering $N_{t}$ being typically large, we assume the hybrid beamforming at the AP with 1-bit DACs to degrade the complexity and infinite-resolution analog-to-digital converters (ADCs) at the user and Eve, due to relatively smaller single antenna.  Moreover, the RIS is equipped with $N_{r}$ reflection elements. In this paper, we adopt a geometric model for mmWave channels \cite{b8}. $\boldsymbol{G}\in\mathbb{C}^{N_{r}\times N_{t}}$ denotes the AP-to-RIS mmWave channel. $\boldsymbol{F}_{RF}\in\mathbb{C}^{N_{t}\times N_{RF}}$ and $\boldsymbol{w}\in\mathbb{C}^{N_{RF}\times 1}$ are the analog beamforming codebook and digital beamforming vector, respectively. $\boldsymbol{F}_{RF}$ adopts the semi-unitary codebook in \cite{b81}.  $\mathcal{Q}(\cdot)$ denotes the 1-bit quantizer.

We define the RIS reflection matrix as $\boldsymbol{\Theta}=\mathrm{diag}(\boldsymbol{\theta})\in\mathbb{C}^{N_{r}\times N_{r}}$, where $\boldsymbol{\theta}=[\beta e^{j\phi_{1}},\ldots, \beta e^{j\phi_{
N_{r}}}]\in\mathbb{C}^{1\times N_{r}}$, $\phi_{i}\in [0,2\pi],~\forall~i=1,\ldots,N_{r}$ and $\beta=1$ denote the phase shifts and the reflection coefficient of the passive element at the RIS, respectively. According to \cite{b13}, due to the hardware limitations, the RIS phase shift $\phi_{i}$ can only choose its phase shifts from a finite number of discrete values $\mathcal{G}=\{0,\Delta\theta,\ldots,(L-1)\Delta\theta\}$, where $L$ is the number of discrete values, and $\Delta\theta=\frac{2\pi}{L}$. Due to unfavorable propagation conditions (obstacles, buildings), the direct link from the AP to the user is ignored.
Then, the RIS reflects the signal to the user and Eve, and the received signals at the user and Eve can be expressed as 
\begin{align}
\textstyle{y=\boldsymbol{h}^{H}\boldsymbol{\Theta}\boldsymbol{G}\boldsymbol{F}_{RF}\mathcal{Q}(\boldsymbol{w}s)+n},
\end{align}
and
\begin{align}
\textstyle{y_{e}=\boldsymbol{h}_{e}^{H}\boldsymbol{\Theta}\boldsymbol{G}\boldsymbol{F}_{RF}\mathcal{Q}(\boldsymbol{w}s)+n_{e}},
\end{align}
in which $n\sim\mathcal{CN}(0,\sigma^{2})$ and $n_{e}\sim\mathcal{CN}(0,\sigma_{e}^{2})$ are the additive white Gaussian noise, $\boldsymbol{h}\in\mathbb{C}^{N_{r}\times 1}$ and $\boldsymbol{h}_{e}\in\mathbb{C}^{N_{r}\times 1}$ are the channel matrix between the RIS and the user, Eve, respectively.

\section{Linear Quantization Models}
We consider linear additive quantization noise model (AQNM) schemes \cite{b91} for the non-linear
quantization operator $\mathcal{Q}(\cdot)$. {The linearization approximation is expressed as
\begin{align}
\textstyle{\mathcal{Q}(\boldsymbol{w}s)\approx b_{Q}\boldsymbol{w}s+\boldsymbol{q}_{Q}},
\end{align}
where $b_{Q}$ is the weight. And it is expressed as
\begin{align}
\textstyle{b_{Q}=1-\eta_{b}},
\end{align}
where $\eta_{b}$ is the distortion factor, which is generally approximated by $\eta_{b}=\frac{\pi\sqrt{3}}{2}2^{-2b}$ for $b$-bit quantization.} The more accurate value for 1-bit quantization is $\eta_{b}\approx 0.3634$ \cite{b91}. In (3), $\boldsymbol{q}_{Q}$ stands for the quantization distortion with the following covariance
\begin{align}
\textstyle{\boldsymbol{A}_{Q}=b_{Q}(1-b_{Q})\mathrm{diag}(\boldsymbol{w}\boldsymbol{w}^{H})}.
\end{align}

Then, the achievable rate at the user can be expressed as
{\begin{align}
&\textstyle{R=\log_{2}\left(1+\frac{|b_{Q}\boldsymbol{h}^{H}\boldsymbol{\Theta}\boldsymbol{G}\boldsymbol{F}_{RF}\boldsymbol{w}|^{2}}{b_{Q}(1-b_{Q})\|\boldsymbol{h}^{H}\boldsymbol{\Theta}\boldsymbol{G}\boldsymbol{F}_{RF}\mathrm{diag}(\boldsymbol{w})\|^{2}+\sigma^{2}}\right)}.
\end{align}
The achievable rate at the Eve can be expressed as
\begin{align}
&\textstyle{R_{e}=\log_{2}\left(1+\frac{|b_{Q}\boldsymbol{h}_{e}^{H}\boldsymbol{\Theta}\boldsymbol{G}\boldsymbol{F}_{RF}\boldsymbol{w}|^{2}}{b_{Q}(1-b_{Q})\|\boldsymbol{h}_{e}^{H}\boldsymbol{\Theta}\boldsymbol{G}\boldsymbol{F}_{RF}\mathrm{diag}(\boldsymbol{w})\|^{2}+\sigma_{e}^{2}}\right)}.
\end{align}}
The system secrecy rate can be written as
\begin{align}
\textstyle{R_{s}=[R-R_{e}]^{+}},
\end{align}
where $[x]^{+}=\max(0,x)$. Then, the problem is formulated as
{\begin{subequations}
\begin{align}
\textstyle{\max\limits_{\boldsymbol{\theta},\boldsymbol{w}}}~&\textstyle{R_{s}}\\
\mbox{s.t.}~
&\textstyle{\textstyle{|\theta_{i}|=1}}&\\
&\textstyle{\|\boldsymbol{F}_{RF}\mathcal{Q}(\boldsymbol{w}s)\|^{2}\leq P},&
\end{align}
\end{subequations}}%
where (9b) denotes the unit-modulus constraint. (9c) is the power constraint, and $P$ is the maximum transmit power. Incorporating (9c) and (5), the transmit power constraint is rewritten as
{\begin{align}
\textstyle{\|\boldsymbol{F}_{RF}\mathcal{Q}(\boldsymbol{w}s)\|^{2}}&\textstyle{=\|b_{Q}\boldsymbol{F}_{RF}\boldsymbol{w}\|^{2}+\mathrm{tr}(\boldsymbol{F}_{RF}\boldsymbol{A}_{Q}\boldsymbol{F}_{RF})},\nonumber\\
&\textstyle{=\|b_{Q}\boldsymbol{w}\|^{2}+\mathrm{tr}(\boldsymbol{A}_{Q})},
\end{align}}%
which follows from the semi-unitary $\boldsymbol{F}_{RF}$. Therefore, problem (9) is rewritten as
{\begin{subequations}
\begin{align}
\textstyle{\max\limits_{\boldsymbol{\Theta},\boldsymbol{w}}}~&\textstyle{R_{s}}\\
\textstyle{\mbox{s.t.}}~
&\textstyle{\text{(9b)},}&\\
&\textstyle{\|b_{Q}\boldsymbol{w}\|^{2}+\mathrm{tr}(\boldsymbol{A}_{Q})\leq P}.&
\end{align}
\end{subequations}}

\section{Alternating Optimization Algorithm}
Since the problem in (11) is non-convex, we adopt an AO-based algorithm to optimize $\boldsymbol{w}$
and $\boldsymbol{\Theta}$, respectively. Specifically, we first optimize $\boldsymbol{w}$ with fixed $\boldsymbol{\Theta}$, then we fix $\boldsymbol{w}$ and optimize $\boldsymbol{\Theta}$.
\subsection{Digital Beamforming Optimization}
Under given $\boldsymbol{\Theta}$, problem (11) is rewritten as
\begin{subequations}
\begin{align}
\textstyle{\max\limits_{\boldsymbol{w}}}~&\textstyle{\log_{2}\left(1+\frac{|\boldsymbol{D}\boldsymbol{w}|^{2}}{\omega}\right)-\log_{2}\left(1+\frac{|\boldsymbol{D}_{e}\boldsymbol{w}|^{2}}{\omega_{e}}\right)}\\
\textstyle{\mbox{s.t.}}~
&\textstyle{\text{(11c)},}&
\end{align}
\end{subequations}
where
\begin{align}
&\textstyle{\boldsymbol{D}=b_{Q}\boldsymbol{h}^{H}\boldsymbol{\Theta}\boldsymbol{G}\boldsymbol{F}_{RF},\boldsymbol{D}_{e}=b_{Q}\boldsymbol{h}^{H}\boldsymbol{\Theta}\boldsymbol{G}\boldsymbol{F}_{RF}},\nonumber\\
&\textstyle{\omega=|\boldsymbol{h}^{H}\boldsymbol{\Theta}\boldsymbol{G}\boldsymbol{F}_{RF}\boldsymbol{A}_{Q}\mathrm{diag}(\boldsymbol{w})|^{2}+\sigma^{2}},\nonumber\\
&\textstyle{\omega_{e}=|\boldsymbol{h}_{e}^{H}\boldsymbol{\Theta}\boldsymbol{G}\boldsymbol{F}_{RF}\boldsymbol{A}_{Q}\mathrm{diag}(\boldsymbol{w})|^{2}+\sigma_{e}^{2}}.
\end{align}
Then, to deal with the non-convex objective function in (12a), we introduce an auxiliary variable $t$ to rewrite (12) as
\begin{subequations}
\begin{align}
\textstyle{\max\limits_{\boldsymbol{w}}}~&\textstyle{\log_{2}\left(1+\frac{|\boldsymbol{D}\boldsymbol{w}|^{2}}{\omega}\right)-t}\\
\mbox{s.t.}~
&\textstyle{\text{(11c)}},&\\
&\textstyle{t=\log_{2}\left(1+\frac{|\boldsymbol{D}_{e}\boldsymbol{w}|^{2}}{\omega_{e}}\right)}.&
\end{align}
\end{subequations}
It is not difficult to find that the objective function is still non-convex. To handle the non-convex objective function, we use $\rho$ to further transform (14a) as
\begin{align}
\textstyle{\max\limits_{\boldsymbol{w},\rho}\log_{2}\left(1+\rho\right)-t}
\end{align}
where
\begin{align}
\textstyle{\rho\leq \frac{|\boldsymbol{D}\boldsymbol{w}|^{2}}{\omega}}.
\end{align}
According to the Schur complement in \cite{b15}, we have
{\begin{align}
\textstyle{\left[
 \begin{matrix}
   1 & z \\
   z & |\boldsymbol{D}\boldsymbol{w}|^{2}
  \end{matrix}
  \right]\succeq\boldsymbol{0}},
\textstyle{~~\rho\leq \frac{z^{2}}{\omega}},
\end{align}
where $z$ is an auxiliary variable.} Due to $\rho\leq \frac{z^{2}}{\omega}$ is non-convex with respect to $\rho$ and $\omega$, we use the SCA method based on the first-order Taylor expansion \cite{b3} to tackle $\rho\leq \frac{z^{2}}{\omega}$. In particular, for any fixed points $(\bar{z},\bar{\omega})$, we have
\begin{align}
\textstyle{\frac{z^{2}}{\omega}\geq \frac{2\bar{z}}{\bar{\omega}}z-\frac{\bar{z}^{2}}{\omega^{2}}\omega\geq \rho}.
\end{align}
By applying the concept of the SCA \cite{b3},\cite{b4}, we iteratively update $\bar{z}$ and $\bar{\omega}$ in the $n$-th iteration as
\begin{align}
\textstyle{\bar{\omega}^{(n)}=\omega^{(n-1)}, 
\bar{z}^{(n)}=z^{(n-1)}}.
\end{align}
Now, we handle the non-convex constraint in (14c), similarly, we use the Schur complement to deal with (14c) and (14c) can be transformed into the following equivalent forms.
\begin{align}
\textstyle{\left[
 \begin{matrix}
   2^{t}-1 & r \\
   r & \omega_{e}
  \end{matrix}
  \right]\succeq\boldsymbol{0}},
\end{align}
\begin{align}
\textstyle{r^{2}\geq (2^{t}-1)\omega_{e}},
\end{align}
and
\begin{align}
\textstyle{r^{2}-|\boldsymbol{D}_{e}\boldsymbol{w}|^{2}=0}.
\end{align}

In order to deal with the bilinear function on the left-hand side of (21), the SCA method based on the arithmetic geometric mean (AGM) inequality is adopted, so (21) can be rewritten as
{}{\begin{align}
\textstyle{\frac{1}{2}\left((\omega_{e}\eta)^{2}+(\frac{2^{t}-1}{\eta})^{2}\right)-\bar{r}(r-\bar{r})\leq 0},   
\end{align}}
where $\eta$ is  a  feasible  point.  To  tighten  the  upper bound, $\eta$ is iteratively updated. The update formulation in the $n$-th iteration is expressed as
\begin{align}
\textstyle{\eta^{(n)}=\sqrt{\frac{2^{t^{(n-1)}}-1}{\rho^{(n-1)}}}}.
\end{align}
Then, we deal with non-convex constraint (22) and use the SCA method to transform (22) as the following convex constraints.
\begin{align}
\textstyle{|\boldsymbol{D}_{e}\boldsymbol{w}|^{2}-\bar{r}(r-\bar{r})\leq 0},~~~~~~~~~~~~~~~~~~~~~~~~~~~\nonumber\\
\textstyle{r^{2}-|\boldsymbol{D}_{e}\bar{\boldsymbol{w}}|^{2}-2\mathrm{Re}(\bar{\boldsymbol{w}}^{H}\boldsymbol{D}_{e}^{H}\boldsymbol{D}_{e}(\boldsymbol{w}-\bar{\boldsymbol{w}}))\leq 0.}
\end{align}
The non-convex constraint in (11c) can be rewritten as the following form by using an similar manner with (25).
\begin{align}
(|b_{Q}\bar{\boldsymbol{w}}|^{2}-2\mathrm{Re}(b_{Q}\bar{\boldsymbol{w}}^{H}(\boldsymbol{w}-\bar{\boldsymbol{w}})))+\mathrm{tr}(\boldsymbol{A_{Q}})\leq P.
\end{align}
Now, the problem in (12) is transformed to the following convex approximation problem.
\begin{subequations}
\begin{align}
\textstyle{\max\limits_{t,r,\omega,\rho,z,\boldsymbol{w}}}~&\textstyle{\log_{2}\left(1+\rho\right)-t}\\
\mbox{s.t.}~
&\left[
 \begin{matrix}
   1 & z \\
   z & |\boldsymbol{D}\boldsymbol{w}|^{2}
  \end{matrix}
  \right]\succeq\boldsymbol{0},\\
&\textstyle{\text{(18)}, \text{(20)}, \text{(23)}, \text{(25)},\text{(26)}.}&
\end{align}
\end{subequations}
The SCA-based algorithm is summarized in \textbf{Algorithm~1} for solving (12). 
\begin{algorithm}[htbp]
\caption{SCA-based algorithm for problem (12).} 
\hspace*{0.02in}{\bf Initialization:} $\bar{r}^{(0)}$, $\bar{\omega}^{(0)}$, $\bar{z}^{(0)}$, $\bar{\boldsymbol{w}}^{(0)}$.\\
\hspace*{0.02in}{\bf Repeat}\\
Update $\{\boldsymbol{w}^{(n)}, r^{(n)}, \rho^{(n)}, \omega^{(n)}, z^{(n)},t^{(n)}\}$ with fixed $\bar{r}$, $\bar{\omega}$, $\bar{t}$, $\bar{\boldsymbol{w}}$ by solving (27).\\
Update $\eta^{(n+1)}$, $\bar{\omega}^{(n+1)}$, $\bar{z}^{(n+1)}$ based on (19) and (24).\\
Update $n=n+1$.\\
\hspace*{0.02in}{\bf Until}
Convergence.\\
\hspace*{0.02in}{\bf Output:} 
$\boldsymbol{w}^{*}$.\\
\end{algorithm}

\subsection{Discrete RIS Phase Shifts Optimization}
Substituting the transmit beamforming vector $\boldsymbol{w}$ obtained in the previous section into problem (11). Then the sub-problem of phase shifts matrix can be rewritten at the top of next page. 

Since the variable $\phi_{i}$ belongs to a finite number of values, problem (28) can be resolved by using the exhaustive search method. However, feasible set $\mathcal{G}$ is large and the complexity of exhaustive search method is very high. To reduce complexity, an element-wise BCD algorithm is proposed.

To obtain the closed form solution of phase shifts, we employ the element-wise BCD algorithm \cite{b3} to solve this problem. We assume $\phi_{i}$ as one block in the BCD and iteratively derive the continuous solutions of $\phi_{i}$ by using \textbf{Theorem~1}.
\setcounter{equation}{28}
\begin{theorem}
There exists solution $\phi_{i}^{*}$ to maximize the value of (28) by 
solving the following equation
\begin{align}
\textstyle{\frac{(\mu_{i}-\bar{\mu}_{i})t-\tilde{\mu}_{i}}{\mu_{i}(1+t^{2})+\bar{\mu}_{i}(1-t^{2})-\tilde{\mu}_{i}2t}-
\frac{(\eta_{i}-\bar{\eta}_{i})t-\tilde{\eta}_{i}}{\eta_{i}(1+t^{2})+\bar{\eta}_{i}(1-t^{2})-\tilde{\eta}_{i}2t}}\nonumber\\
\textstyle{+\frac{(\lambda_{i}-\bar{\lambda}_{i})t-\tilde{\lambda}_{i}}{\lambda_{i}(1+t^{2})+\bar{\lambda}_{i}(1-t^{2})-\tilde{\lambda}_{i}2t}-
\frac{(\rho_{i}-\bar{\rho}_{i})t-\tilde{\rho}_{i}}{\rho_{i}(1+t^{2})+\bar{\rho}_{i}(1-t^{2})-\tilde{\rho}_{i}2t}=0}.
\end{align}
We use one dimension search method in \cite{b101} to solve the equation in (29). $\phi_{i}^{*}$ can be expressed as
\begin{eqnarray}
\phi_{i}^{*}=2\arctan(t).
\end{eqnarray}

The proof is given in \textbf{Appendix~A}.
\end{theorem}
\setcounter{equation}{30}
According to \textbf{Theorem~1}, the discrete solution
$\bar{\phi}_{i}^{*}$ can be calculated by the following formulation
\begin{align}
\bar{\phi}_{i}^{*}=\arg\min_{\phi_{i}\in\mathcal{G}}|\phi_{i}^{*}-\phi_{i}|.
\end{align}
The algorithm based on element-wise BCD can be summarized in \textbf{Algorithm~2}.

\newcounter{mytempeqncnt}
\begin{figure*}[!t]
\normalsize
\setcounter{mytempeqncnt}{\value{equation}}
\setcounter{equation}{27}
{\begin{subequations}
\begin{align}
\textstyle{\max\limits_{\boldsymbol{\Theta}}}~&\textstyle{\frac{\left(\frac{b_{Q}(1-b_{Q})\|\boldsymbol{\theta}^{T}\mathrm{diag}(\boldsymbol{h}^{H})\boldsymbol{G}\boldsymbol{F}_{RF}\mathrm{diag}(\boldsymbol{w})\|^{2}+\sigma^{2}+|b_{Q}\boldsymbol{h}^{H}\boldsymbol{\Theta}\boldsymbol{G}\boldsymbol{F}_{RF}\boldsymbol{w}|^{2}}{b_{Q}(1-b_{Q})\|\boldsymbol{\theta}^{T}\mathrm{diag}(\boldsymbol{h}^{H})\boldsymbol{G}\boldsymbol{F}_{RF}\mathrm{diag}(\boldsymbol{w})\|^{2}+\sigma^{2}}\right)}{\left(\frac{b_{Q}(1-b_{Q})\|\boldsymbol{\theta}^{T}\mathrm{diag}(\boldsymbol{h}_{e}^{H})\boldsymbol{G}\boldsymbol{F}_{RF}\mathrm{diag}(\boldsymbol{w})\|^{2}+\sigma^{2}+|b_{Q}\boldsymbol{h}_{e}^{H}\boldsymbol{\Theta}\boldsymbol{G}\boldsymbol{F}_{RF}\boldsymbol{w}|^{2}}{b_{Q}(1-b_{Q})\|\boldsymbol{\theta}^{T}\mathrm{diag}(\boldsymbol{h}_{e}^{H})\boldsymbol{G}\boldsymbol{F}_{RF}\mathrm{diag}(\boldsymbol{w})\|^{2}+\sigma_{e}^{2}}\right)}}&\\
\mbox{s.t.}~
&\text{(9b)}&
\end{align}
\end{subequations}}
\setcounter{equation}{\value{mytempeqncnt}}
\hrulefill
\vspace*{4pt}
\end{figure*}
\begin{algorithm}[htbp]
\caption{Element-wise BCD-based algorithm for problem (28).} 
\hspace*{0.02in}{\bf Initialization:} $t=0$, $\boldsymbol{\Theta}^{0}$.\\
\hspace*{0.02in}{\bf Repeat:}\\
\hspace*{0.02in}{\bf for:}~$i=1,\ldots,N_{r}$\\
Calculate $\phi_{i}^{t}$ based on \textbf{Theorem~1}.
\\
\hspace*{0.02in}{\bf End:}\\
Set $t=t+1$\\
\hspace*{0.02in}{\bf Until:}
Convergence.\\
\hspace*{0.02in}{\bf Output:}
$\boldsymbol{\Theta}^{*}$
\end{algorithm}
Based on the above analysis, we can obtain the AO-based algorithm for solving problem (9). Following the results in \cite{b3}, since each sub-algorithm converges to a local optimal solution,  we can guarantee that the AO-based algorithm can converge to a local optimal solution.
\begin{algorithm}[htbp]
\caption{AO-based Algorithm for problem (9).} 
\hspace*{0.02in}{\bf Initialization:} $\boldsymbol{w}^{(0)}$, $\boldsymbol{\Theta}^{(0)}$.\\
\hspace*{0.02in}{\bf Repeat}\\
Update $\boldsymbol{w}^{(j)}$ by using \textbf{Algorithm~1}.\\
Update $\boldsymbol{\Theta}^{(j)}$ by using \textbf{Algorithm~2}.\\
Update $j=j+1$.\\
\hspace*{0.02in}{\bf Until}
Convergence.\\
\hspace*{0.02in}{\bf Output:} 
$\boldsymbol{w}^{*}$, $\boldsymbol{\Theta}^{*}$.\\
\end{algorithm}

\subsection{Complexity Analysis}
 The complexity of the proposed method is about $\mathcal{O}(N_{r}(N_{r}N_{t}+L_{P})+N_{t}^{2})$, which depends on the computational complexity of $\phi_{i}$ and $\boldsymbol{w}$. We compare the complexity of these algorithms in \textbf{TABLE~1}, it is not difficult to find that the AO-based algorithm has the lowest complexity.
 \begin{table}[htbp]
  \centering
  \scriptsize
  \caption{Complexity Comparison of Algorithms }
  \label{tab:notations}
  \begin{tabular}{ll}
    \\[-2mm]
    \hline
    \hline\\[-2mm]
    {\bf \small Symbol}&\qquad {\bf\small Total Complexity}\\
    \hline
    \vspace{0.7mm}\\[-2mm]
    Proposed AO-based algorithm      &   $\mathcal{O}(N_{r}(N_{r}N_{t}+L_{P})+N_{t}^{2})$\\
    \vspace{0.7mm}
    the exhaustive method         &  $\mathcal{O}(N_{r}^{L_{P}+1} +N_{r}^{2}+N_{r}N_{t})$\\
     \vspace{0.7mm}
    SDP-based algorithm~\cite{b100}          &  $\mathcal{O}(\zeta N_{r}^{8}+N_{t}^{2})$\\
    \hline
    \hline
  \end{tabular}
\end{table}

\section{Numerical Results}\label{IV}
As shown in Fig.1, we consider a RIS-aided mmWave system with hardware limitations, where AP's coordinate is $(0,0,0)$~m and IRS is located at $(0, 60, 20)$~m.  While the user and the Eve are located at $(5,60,0)$~m and $(5,80,0)$~m, respectively. {We set $N_{t}=64$, $N_{r}=16$, $N_{RF}=8$, $b=1$, $\sigma^{2}=-110$~dBm.} 
The mmWave channels from the AP to the RIS, the RIS to the $k$th user are respectively expressed as 
\begin{eqnarray}
\textstyle{\boldsymbol{G}=\sqrt{\frac{1}{\beta L_{1}}}\sum_{l=0}^{L_{1}-1}\alpha\boldsymbol{a}_{T}(N_{T},\theta)\boldsymbol{a}_{R}^{T}(N_{r},\varphi,\phi),}
\end{eqnarray}
\begin{eqnarray}
\textstyle{\boldsymbol{h}_{k}=\sqrt{\frac{1}{\hat{\beta}_{k}L_{k}}}\sum_{l=0}^{L_{k}-1}\hat{\alpha}_{k}\boldsymbol{a}_{R}(N_{r},\vartheta_{k}),~k\in\{user, Eve\},}
\end{eqnarray}
where $\beta$ and $\hat{\beta}_{k}$ denote the large-scale fading coefficients. They are generated by
\begin{align}
72+29.2\log_{10}d+\zeta,
\end{align}
where $d$ denotes the signal propagation distance, $\zeta\sim\mathcal{CN}(0,1)$ is the log-normal shadowing, $\alpha$ and $\hat{\alpha}_{k}$ denote the small-scale fading coefficients which follow $\mathcal{CN}(0,1)$ [15]. $\boldsymbol{a}_{T}(\cdot)$ and $\boldsymbol{a}_{R}(\cdot)$ represent array steering vectors at the AP and the RIS, respectively.
Three schemes with fixed quantization bits for each DACs are considered for comparison:
\begin{itemize}
\item MRT-BCD: In this scheme, the transmit beamforming is designed based on maximum ratio transmission (MRT). Then, the RIS phase shifts is obtained by using BCD algorithm.
\item NO-RIS: In this scheme, the mmWave do not use the RIS to assist communication. Moreover, the transmit beamforming is optimized by using \textbf{Algorithm~1}.
\item Upper Bound: In this scheme, we consider the RIS-aided mmWave system without hardware limitations as an upper bound. Moreover, the transmit beamforming vector and the RIS phase shifts are optimized by using \textbf{Algorithm~1} and \textbf{Algorithm~2}, respectively.
\end{itemize}

\begin{figure*}[t]
\centering
\subfigure{
\begin{minipage}[t]{0.35\linewidth}
\centering
\includegraphics[scale=0.3]{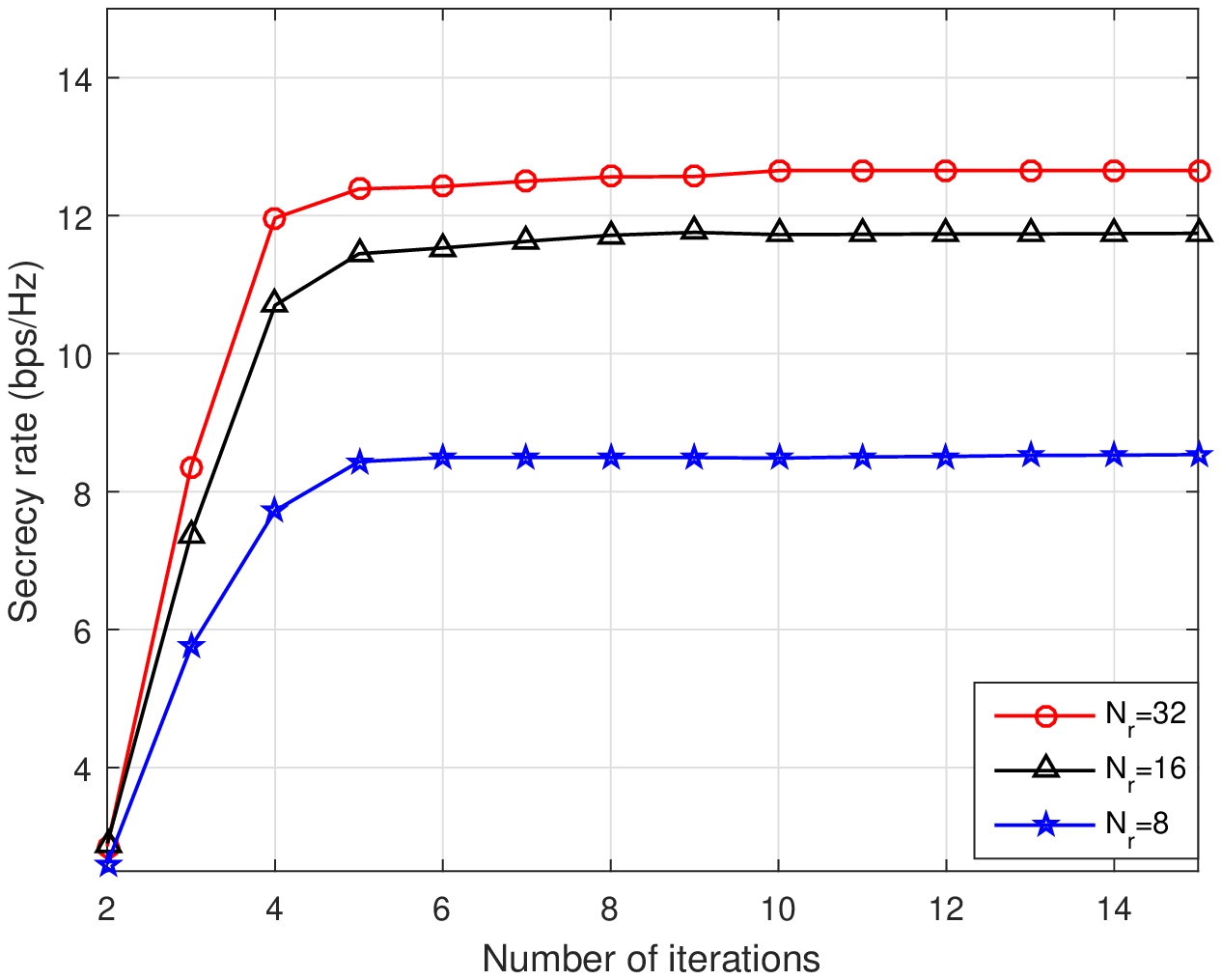}
\caption{Convergence of \textbf{Algorithm~3}}
\end{minipage}%
}%
\subfigure{
\begin{minipage}[t]{0.35\linewidth}
\centering
\includegraphics[scale=0.3]{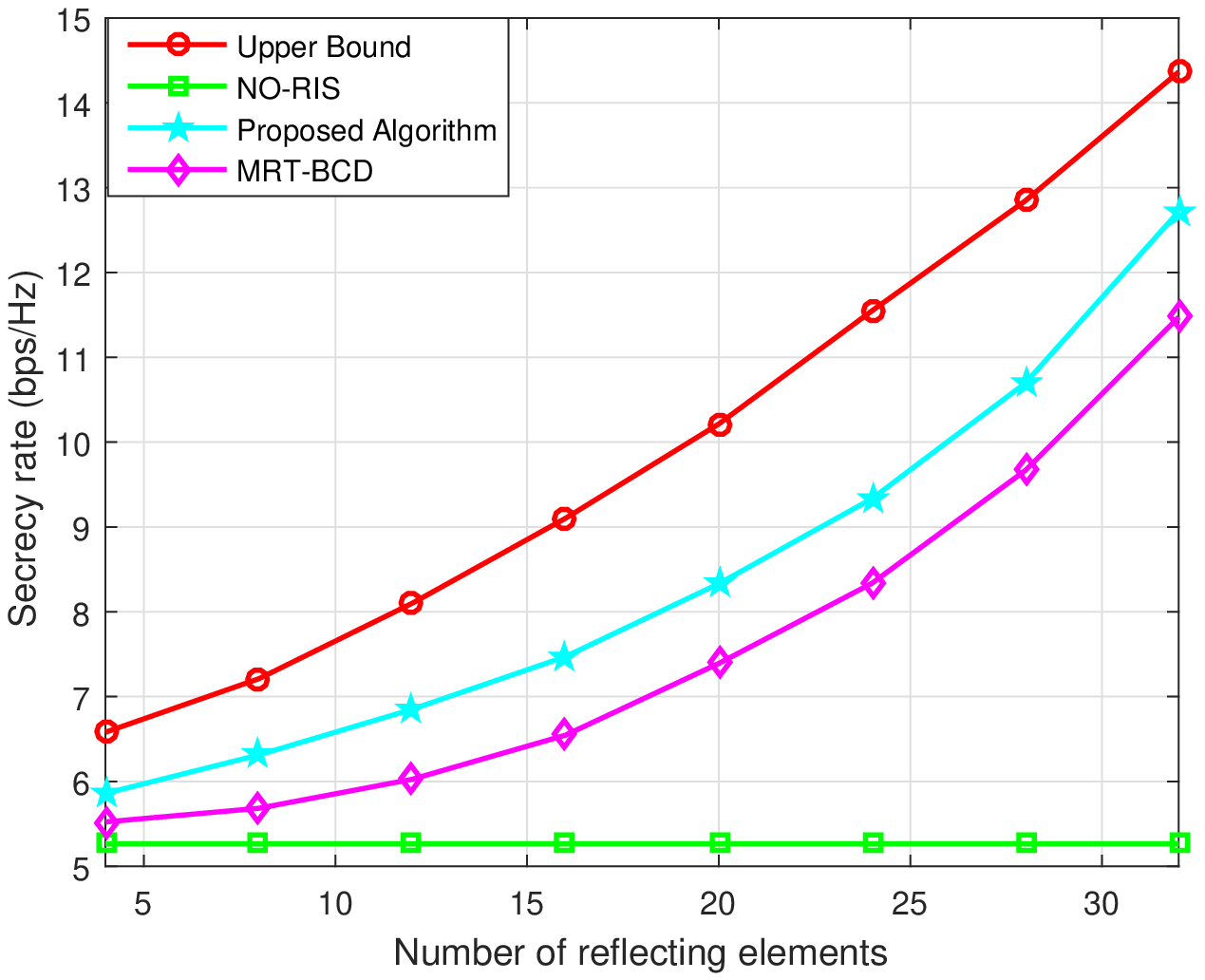}
\caption{Secrecy rate versus $N_{r}$}
\end{minipage}%
}%
\subfigure{
\begin{minipage}[t]{0.35\linewidth}
\centering
\includegraphics[scale=0.30]{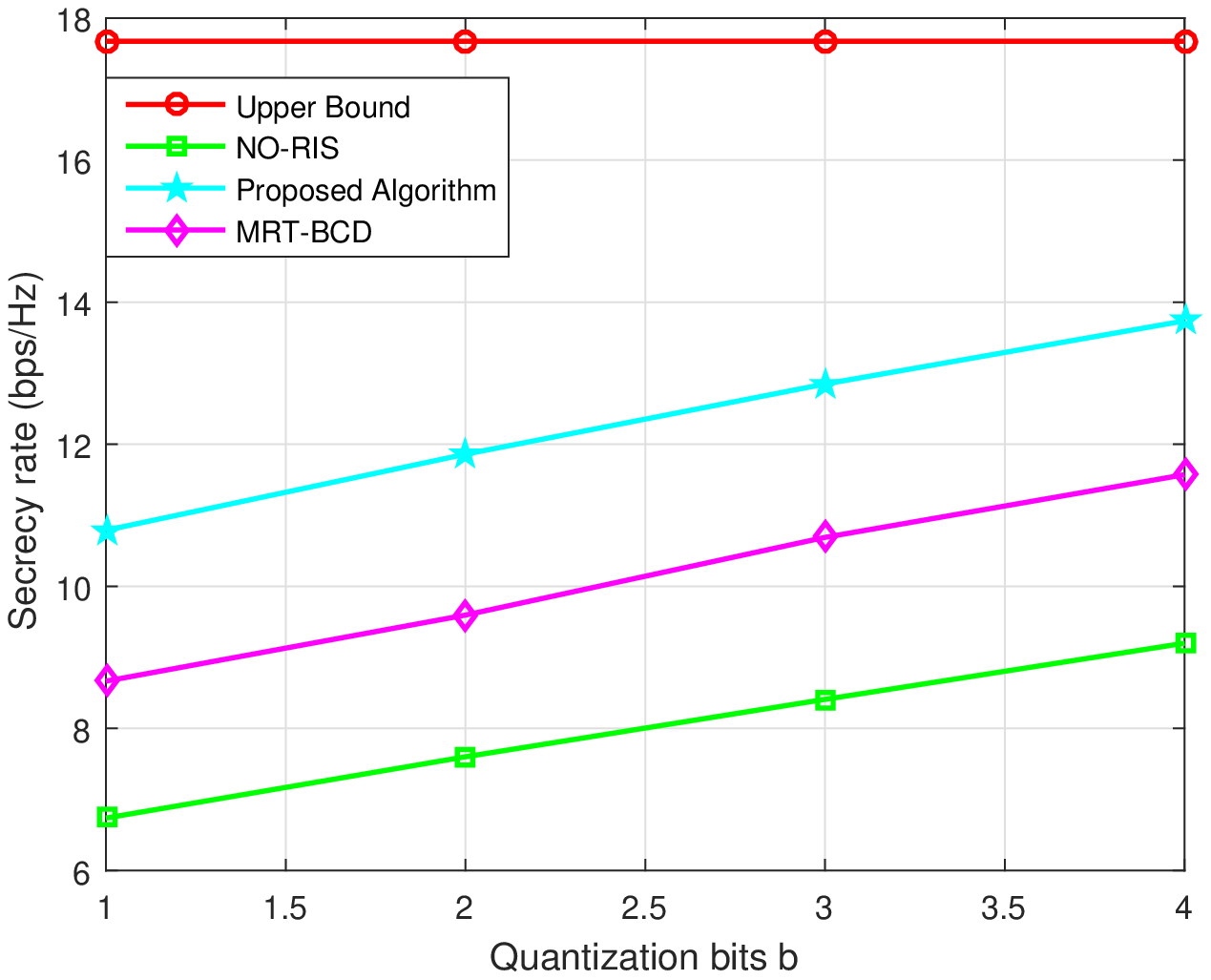}
\caption{Secrecy rate versus $b$}
\end{minipage}
}%
\centering
\end{figure*}
{The convergence behavior of the proposed AO-based algorithm under different iteration times is given in Fig.2. We observe that the secrecy rate increases monotonically with the increase of the number of iterations. In addition, the algorithm converges quickly, and can achieve a high security rate at the 5th iteration. }

{Fig. 3 shows the secrecy rate under different number of reflecting elements of the RIS. 
As expected, compared with the MRT-BCD and NO-RIS schemes, the proposed algorithm can achieve the best performance. 
Moreover, although the resolution of DACs in the upper bound scheme is higher than that of the proposed scheme, we find that as the number of reflecting elements of RIS increases, both the proposed scheme and the upper bound scheme increase at the same time. This finding validates the feasibility of using RIS mitigates the influence of hardware limitations.}

{Fig. 4 depicts the secrecy rate versus quantization bits under different schemes. Compared with the MRT-BCD scheme and NO-RIS scheme, the proposed AO-based algorithm can achieve the best security performance.  
It is not difficult to find that the schemes with the RIS can outperformance the NO-RIS scheme. From the perspective of maximizing the secrecy rate, when the system is assisted by RIS, there is no need to deploy high-resolution DACs. This result demonstrates that the RIS can suppress the hardware loss as well.}

\section{Conclusion}
The secrecy rate maximization problem for mmWave communications with hardware limitations at the AP and RIS was investigated in this paper. The RIS phase shifts, transmit beamforming vector were jointly optimized to maximize the secrecy rate under the hardware constraint and unit-modulus constraints. To solve this problem, we proposed an AO-based algorithm. Numerical results have shown that the proposed AO-based algorithm outperforms conventional schemes in terms of secrecy rate. Moreover, the numerical results also show that when the mmWave system is equipped with RIS, the mmWave system does not need to equip with excessive RF chains and high-resolution DACs. {In this paper, the CSI is assumed as perfect. In fact, the CSI is difficult to obtain due to the passive feature of the RIS. Therefore, we will design a channel estimation scheme
\footnote{{We carefully studied the channel estimation scheme in \cite{b111,b222}. They are useful to improve our future work, and we will combine the channel estimation results to design the beamforming for RIS-aided mmWave system.}}in the future works, then we use the proposed algorithm in this paper to design beamforming for maximizing secrecy rate in a RIS-aided mmWave system.}

\appendices
\section{The proof of \textbf{theorem~1}}
We assume $\phi_{i}$ as one block of BCD, (28a) can be rewritten as
\begin{align}
\textstyle{R_{s}=\left(\frac{|\boldsymbol{\theta}^{T}\boldsymbol{c}|^{2}+\|\boldsymbol{\theta}^{T}\boldsymbol{A}\|^{2}+\sigma^{2}}{|\boldsymbol{\theta}^{T}\boldsymbol{d}|^{2}+\|\boldsymbol{\theta}^{T}\boldsymbol{B}\|^{2}+\sigma_{e}^{2}}\right)\left(\frac{\|\boldsymbol{\theta}^{T}\boldsymbol{B}\|^{2}+\sigma_{e}^{2}}{\|\boldsymbol{\theta}^{T}\boldsymbol{A}\|^{2}+\sigma^{2}}\right)},
\end{align}
where
\begin{align}
&\textstyle{\boldsymbol{c}=b_{Q}\mathrm{diag}(\boldsymbol{h}^{H})\boldsymbol{G}\boldsymbol{F}_{RF}\boldsymbol{w}, \boldsymbol{d}=b_{Q}\mathrm{diag}(\boldsymbol{h}_{e}^{H})\boldsymbol{G}\boldsymbol{F}_{RF}\boldsymbol{w}},\nonumber\\
&\textstyle{\boldsymbol{A}=b_{Q}(1-b_{Q})\mathrm{diag}(\boldsymbol{h}^{H})\boldsymbol{G}\boldsymbol{F}_{RF}\mathrm{diag}(\boldsymbol{w})},\nonumber\\
&\textstyle{\boldsymbol{B}=b_{Q}(1-b_{Q})\mathrm{diag}(\boldsymbol{h}_{e}^{H})\boldsymbol{G}\boldsymbol{F}_{RF}\mathrm{diag}(\boldsymbol{w})}.
\end{align}
To simplify (35), we expanse $|\boldsymbol{\theta}^{T}\boldsymbol{c}|^{2}$, $|\boldsymbol{\theta}^{T}\boldsymbol{d}|^{2}$, $|\boldsymbol{\theta}^{T}\boldsymbol{A}|^{2}$, and $|\boldsymbol{\theta}^{T}\boldsymbol{B}|^{2}$ as
\begin{align}
&\textstyle{|\boldsymbol{\theta}^{T}\boldsymbol{c}|^{2}=|e^{j\phi_{i}}c_{i}+p_{i}|^{2}=|c_{i}|^{2}+|p_{i}|^{2}}\nonumber\\
&\textstyle{+(\mathrm{Re}\{c_{i}\}\mathrm{Re}\{p_{i}\}+\mathrm{Im}\{c_{i}\}\mathrm{Im}\{p_{i}\})\cos(\phi_{i})}\nonumber\\
&\textstyle{-(\mathrm{Re}\{c_{i}\}\mathrm{Im}\{p_{i}\}+\mathrm{Im}\{c_{i}\}\mathrm{Re}\{p_{i}\})\sin(\phi_{i})},\nonumber\\
&\textstyle{|\boldsymbol{\theta}^{T}\boldsymbol{d}|^{2}=|e^{j\phi_{i}}\bar{c}_{i}+\bar{p}_{i}|^{2}=|\bar{c}_{i}|^{2}+|\bar{p}_{i}|^{2}}\nonumber\\
&\textstyle{+(\mathrm{Re}\{\bar{c}_{i}\}\mathrm{Re}\{\bar{p}_{i}\}+\mathrm{Im}\{\bar{c}_{i}\}\mathrm{Im}\{\bar{p}_{i}\})\cos(\phi_{i})}\nonumber\\
&\textstyle{-(\mathrm{Re}\{\bar{c}_{i}\}\mathrm{Im}\{\bar{p}_{i}\}+\mathrm{Im}\{\bar{c}_{i}\}\mathrm{Re}\{\bar{p}_{i}\})\sin(\phi_{i})},\nonumber\\
&\textstyle{|\boldsymbol{\theta}^{T}\boldsymbol{A}|^{2}=\sum_{k=1}^{N_{t}}|e^{j\phi_{i}}q_{i,k}+v_{k}|^{2}=\sum_{k=1}^{N_{t}}|q_{i,k}|^{2}+\sum_{k=1}^{N_{t}}|v_{k}|^{2}}\nonumber\\
&\textstyle{+\sum_{k=1}^{N_{t}}(\mathrm{Re}\{q_{i,k}\}\mathrm{Re}\{v_{k}\}+\mathrm{Im}\{q_{i,k}\}\mathrm{Im}\{v_{k}\})\cos(\phi_{i})}\nonumber\\
&\textstyle{-\sum_{k=1}^{N_{t}}(\mathrm{Re}\{q_{i,k}\}\mathrm{Im}\{v_{k}\}+\mathrm{Im}\{q_{i,k}\}\mathrm{Re}\{v_{k}\})\sin(\phi_{i})},\nonumber\\
&\textstyle{|\boldsymbol{\theta}^{T}\boldsymbol{B}|^{2}=\sum_{k=1}^{N_{t}}|e^{j\phi_{i}}\bar{q}_{i,k}+\bar{v}_{k}|^{2}=\sum_{k=1}^{N_{t}}|\bar{q}_{i,k}|^{2}+\sum_{k=1}^{N_{t}}|\bar{v}_{k}|^{2}}\nonumber\\
&\textstyle{+\sum_{k=1}^{N_{t}}(\mathrm{Re}\{\bar{q}_{i,k}\}\mathrm{Re}\{\bar{v}_{k}\}+\mathrm{Im}\{\bar{q}_{i,k}\}\mathrm{Im}\{\bar{v}_{k}\})\cos(\phi_{i})}\nonumber\\
&\textstyle{-\sum_{k=1}^{N_{t}}(\mathrm{Re}\{\bar{q}_{i,k}\}\mathrm{Im}\{\bar{v}_{k}\}+\mathrm{Im}\{\bar{q}_{i,k}\}\mathrm{Re}\{\bar{v}_{k}\})\sin(\phi_{i})},
\end{align}
where
\begin{align}
&\textstyle{p_{i}=\sum_{j\neq i}^{N_{r}}e^{j\phi_{j}}c_{j},     \bar{p}_{i}=\sum_{j\neq i}^{N_{r}}e^{j\phi_{j}}d_{j},\bar{c}_{i}=d_{i}},\nonumber\\
&\textstyle{q_{i,k}=a_{i,k}, \bar{q}_{i,k}=b_{i,k}}\nonumber\\
&\textstyle{v_{k}=\sum_{j\neq 1}^{N_{r}}e^{j\phi_{j}}a_{k,j},     \bar{v}_{k}=\sum_{j\neq 1}^{N_{r}}e^{j\phi_{j}}b_{k,j}}
\end{align}
We continue to simplify (35) by introducing auxiliary variables $\mu_{i}$, $\bar{\mu}$, $\eta_{i}$, $\bar{\eta}_{i}$, $\lambda_{i}$, $\bar{\lambda}_{i}$, $\rho_{i}$, and $\bar{\rho}_{i}$.
\begin{align}
\textstyle{R_{s}=\left(\frac{\mu_{i}+\bar{\mu}_{i}\cos(\phi_{i})-\tilde{\mu}_{i}\sin(\phi_{i})}{\eta_{i}+\bar{\eta}_{i}\cos(\phi_{i})-\tilde{\eta}_{i}\sin(\phi_{i})}\right)\left(\frac{\lambda_{i}+\bar{\lambda}_{i}\cos(\phi_{i})-\tilde{\lambda}_{i}\sin(\phi_{i})}{\rho_{i}+\bar{\rho}_{i}\cos(\phi_{i})-\tilde{\rho}_{i}\sin(\phi_{i})}\right)},
\end{align}
where
\begin{align}
&\textstyle{\mu_{i}=|c_{i}|^{2}+|p_{i}|^{2}+\sum_{k=1}^{N_{t}}|q_{i,k}|^{2}+\sum_{k=1}^{N_{t}}|v_{k}|^{2}+\sigma^{2}},\nonumber\\
&\textstyle{\bar{\mu}_{i}=(\mathrm{Re}\{c_{i}\}\mathrm{Re}\{p_{i}\}+\mathrm{Im}\{c_{i}\}\mathrm{Im}\{p_{i}\})}\nonumber\\
&\textstyle{+\sum_{k=1}^{N_{t}}(\mathrm{Re}\{q_{i,k}\}\mathrm{Re}\{v_{k}\}+\mathrm{Im}\{q_{i,k}\}\mathrm{Im}\{v_{k}\})},\nonumber\\
&\textstyle{\tilde{\mu}_{i}=\sum_{k=1}^{N_{t}}(\mathrm{Re}\{q_{i,k}\}\mathrm{Im}\{v_{k}\}+\mathrm{Im}\{q_{i,k}\}\mathrm{Re}\{v_{k}\})},\nonumber\\
&\textstyle{\eta_{i}=|\bar{c}_{i}|^{2}+|\bar{p}_{i}|^{2}+\sum_{k=1}^{N_{t}}|\bar{q}_{i,k}|^{2}+\sum_{k=1}^{N_{t}}|\bar{v}_{k}|^{2}+\sigma^{2}},\nonumber\\
&\textstyle{\bar{\eta}_{i}=(\mathrm{Re}\{\bar{c}_{i}\}\mathrm{Re}\{\bar{p}_{i}\}+\mathrm{Im}\{\bar{c}_{i}\}\mathrm{Im}\{\bar{p}_{i}\})}\nonumber\\
&\textstyle{+\sum_{k=1}^{N_{t}}(\mathrm{Re}\{\bar{q}_{i,k}\}\mathrm{Re}\{\bar{v}_{k}\}+\mathrm{Im}\{\bar{q}_{i,k}\}\mathrm{Im}\{\bar{v}_{k}\})},\nonumber\\
&\textstyle{\tilde{\eta}_{i}=\sum_{k=1}^{N_{t}}(\mathrm{Re}\{\bar{q}_{i,k}\}\mathrm{Im}\{\bar{v}_{k}\}+\mathrm{Im}\{\bar{q}_{i,k}\}\mathrm{Re}\{\bar{v}_{k}\})},\nonumber\\
&\textstyle{\rho_{i}=\sum_{k=1}^{N_{t}}|q_{i,k}|^{2}+\sum_{k=1}^{N_{t}}|v_{k}|^{2}+\sigma^{2}},\nonumber\\
&\textstyle{\bar{\rho}_{i}=\sum_{k=1}^{N_{t}}(\mathrm{Re}\{q_{i,k}\}\mathrm{Re}\{v_{k}\}+\mathrm{Im}\{q_{i,k}\}\mathrm{Im}\{v_{k}\})},\nonumber\\
&\textstyle{\tilde{\rho}_{i}=\sum_{k=1}^{N_{t}}(\mathrm{Re}\{q_{i,k}\}\mathrm{Im}\{v_{k}\}+\mathrm{Im}\{q_{i,k}\}\mathrm{Re}\{v_{k}\}},\nonumber\\
&\textstyle{\lambda_{i}=\sum_{k=1}^{N_{t}}|\bar{q}_{i,k}|^{2}+\sum_{k=1}^{N_{t}}|\bar{v}_{k}|^{2}+\sigma^{2}},\nonumber\\
&\textstyle{\bar{\lambda}_{i}=\sum_{k=1}^{N_{t}}(\mathrm{Re}\{\bar{q}_{i,k}\}\mathrm{Re}\{\bar{v}_{k}\}+\mathrm{Im}\{\bar{q}_{i,k}\}\mathrm{Im}\{\bar{v}_{k}\})},\nonumber\\
&\textstyle{\tilde{\lambda}_{i}=\sum_{k=1}^{N_{t}}(\mathrm{Re}\{\bar{q}_{i,k}\}\mathrm{Im}\{\bar{v}_{k}\}+\mathrm{Im}\{\bar{q}_{i,k}\}\mathrm{Re}\{\bar{v}_{k}\}}.\nonumber\\
\end{align}
Let $\tan(\frac{\phi_{i}}{2})=t$, we have $\sin(\phi_{i})=\frac{2t}{1+t^{2}}$ and $\cos(\phi_{i})=\frac{1-t^{2}}{1+t^{2}}$. 
(39) is rewritten as
\begin{align}
&\textstyle{R_{s}=\left(\frac{\mu_{i}(1+t^{2})+\bar{\mu}_{i}(1-t^{2})-\tilde{\mu}_{i}2t}{\eta_{i}(1+t^{2})+\bar{\eta}_{i}(1-t^{2})-\tilde{\eta}_{i}2t}\right)}\textstyle{\left(\frac{\lambda_{i}(1+t^{2})+\bar{\lambda}_{i}(1-t^{2})-\tilde{\lambda}_{i}2t}{\rho_{i}(1+t^{2})+\bar{\rho}_{i}(1-t^{2})-\tilde{\rho}_{i}2t}\right)}.
\end{align}
When the derivative of (41) is $0$, the optimal solution $\phi_{i}^{*}$ can be determined by the following equation
\begin{align}
&\textstyle{\frac{(\mu_{i}-\bar{\mu}_{i})t-\tilde{\mu}_{i}}{\mu_{i}(1+t^{2})+\bar{\mu}_{i}(1-t^{2})-\tilde{\mu}_{i}2t}-
\frac{(\eta_{i}-\bar{\eta}_{i})t-\tilde{\eta}_{i}}{\eta_{i}(1+t^{2})+\bar{\eta}_{i}(1-t^{2})-\tilde{\eta}_{i}2t}}\nonumber\\
&\textstyle{+\frac{(\lambda_{i}-\bar{\lambda}_{i})t-\tilde{\lambda}_{i}}{\lambda_{i}(1+t^{2})+\bar{\lambda}_{i}(1-t^{2})-\tilde{\lambda}_{i}2t}-
\frac{(\rho_{i}-\bar{\rho}_{i})t-\tilde{\rho}_{i}}{\rho_{i}(1+t^{2})+\bar{\rho}_{i}(1-t^{2})-\tilde{\rho}_{i}2t}=0},
\end{align}\par
Then, the equation (42) can be solved using a one
dimension search in \cite{b101}. Finally, the \textbf{Theorem~1} is proved. 
\ifCLASSOPTIONcaptionsoff
  \newpage
\fi


\bibliographystyle{IEEEtran}

\end{document}